\documentclass[journal]{IEEEtran}
\usepackage{xcolor,soul,framed} %,caption

\colorlet{shadecolor}{yellow}
\usepackage[pdftex]{graphicx}
\graphicspath{{../pdf/}{../jpeg/}}
\DeclareGraphicsExtensions{.pdf,.jpeg,.png}

\usepackage[cmex10]{amsmath}
\usepackage{array}
\usepackage{mdwmath}
\usepackage{mdwtab}
\usepackage{eqparbox}
\usepackage{url}
\usepackage{multirow}
\usepackage{booktabs}
\usepackage{caption}
\usepackage{subcaption}
\usepackage[linesnumbered, ruled,vlined]{algorithm2e}
\usepackage{hyperref}

\graphicspath{{./}}

\begin{document}
    \title{Epileptic Seizure Classification \\ with Symmetric and Hybrid Bilinear Models}
    \author{Tennison~Liu, Nhan~Duy~Truong,~\IEEEmembership{Student Member,~IEEE}, Armin~Nikpour, Luping~Zhou,~\IEEEmembership{Senior Member, IEEE},
    and~Omid~Kavehei$^*$,~\IEEEmembership{Senior Member,~IEEE\thanks{$^*$~Corresponding author.}}
\thanks{T.~Liu, N.D.~Truong, L.~Zhou and O.~Kavehei are with the Faculty of Engineering, The University of Sydney, NSW 2006, Australia.\protect\\
A.~Nikpour is with Department of Neurology at the Royal Prince Alfred Hospital, NSW 2050, and Sydney Medical School, The University of Sydney, NSW 2006, Australia.\protect\\ E-mails: tliu0311@uni.sydney.edu.au,\{duy.truong,luping.zhou,omid.kavehei, armin.nikpour\}@sydney.edu.au}
} 

\maketitle

\begin{abstract}
Epilepsy affects nearly $\mathbf{1\%}$ of the global population, of which two thirds can be treated by anti-epileptic drugs and a much lower percentage by surgery. Diagnostic procedures for epilepsy and monitoring are highly specialized and labour-intensive. The accuracy of the diagnosis is also complicated by overlapping medical symptoms, varying levels of experience and inter-observer variability among clinical professions. This paper proposes a novel hybrid bilinear deep learning network with an application in the clinical procedures of epilepsy classification diagnosis, where the use of surface electroencephalogram (sEEG) and audiovisual monitoring is standard practice. Hybrid bilinear models based on two types of feature extractors, namely Convolutional Neural Networks (CNNs) and Recurrent Neural Networks (RNNs), are trained using Short-Time Fourier Transform (STFT) of one-second sEEG. In the proposed hybrid models, CNNs extract spatio-temporal patterns, while RNNs focus on the characteristics of temporal dynamics in relatively longer intervals given the same input data. Second-order features, based on interactions between these spatio-temporal features are further explored by bilinear pooling and used for epilepsy classification. Our proposed methods obtain an F1-score of $\mathbf{97.4\%}$ on the Temple University Hospital Seizure Corpus and $\mathbf{97.2\%}$ on the EPILEPSIAE dataset, comparing favourably to existing benchmarks for sEEG-based seizure type classification. The open-source implementation of this study is available at \mbox{\small \url{https://github.com/NeuroSyd/Epileptic-Seizure-Classification}}.
\end{abstract}

\begin{IEEEkeywords}
epilepsy, epileptic seizure classification, bilinear models, EEG, deep learning
\end{IEEEkeywords}

\section{Introduction}

\IEEEPARstart{T}he International League Against Epilepsy (ILAE), defines epilepsy as {\lq\lq}a disorder of the brain characterized by an enduring predisposition to generate epileptic seizures{\rq\rq} \cite{fisher2005epileptic,fisher2017operational}. Epilepsy attacks come in different types and are treated differently. Critical treatment and prognosis procedures all rely and start with the correct identification of epileptic seizure type. ILAE classifies seizure types based on different types of diagnosis which includes origin and symptoms \cite{fisher2017operational}. The classification of seizure type is made primarily on clinical grounds based on demographic and medical history variables and is supported by EEG and radiographic studies. Long-term electroencephalogram monitoring with video recording (video-EEG) is the most common supporting method of seizure classification \cite{friedman2009long}. Generally, epilepsy can be successfully treated with anti-epileptic medication. Around $60$-$70\%$ of people diagnosed with seizure will gain seizure control with medication. Surgery is another viable medical option for certain conditions of epileptic seizures. The correct seizure type diagnosis is vital in order to select the appropriate drug therapy and to provide information regarding the prognosis \cite{goldenberg2010overview}. This paper focuses on deep learning tools for automatic epileptic seizure type classification.

\subsection{Diagnostic Challenges in Seizure Classification}
An accurate clinical diagnosis requires a thorough history from the patient and observers, which can be compromised by inaccurate and inadequate patient and witness history \cite{syed2011can}. Overlapping clinical features also plays a contributory role in incorrect differentials as focal and generalized seizure disorders show overlap of both clinical and EEG symptoms \cite{panayiotopoulos2005clinical}. Recent studies have shown that focal and generalized epilepsy are often difficult to differentiate even by experienced neurologists \cite{panayiotopoulos2005optimal}. Clinical diagnosis is further complicated by the variable expression of epilepsy since the manifestation of same classes of the epilepsy can be quite varied between different patients and also for an individual patient over time \cite{panayiotopoulos2005clinical}.

When a diagnosis cannot be reliably reached on clinical grounds, the video-EEG has been shown to be indispensable to confirm epilepsy diagnosis. In many conditions, including infantile spasms, myoclonic epilepsy and idiopathic generalized epilepsy, the video-EEG may specifically confirm or support a correct diagnosis \cite{panayiotopoulos2005optimal}. Video-EEG monitoring involves patients staying in epilepsy monitoring units for several days where natural or induced seizure events are recorded. Neurologists then visually examine these video-EEG records, resulting in a tedious and time-consuming process particularly where hour's or day's worth of EEG needs to be reviewed visually. In many countries, there is a shortage of neurologists and other EEG professionals - manual examination of EEG records delays the diagnosis process and occupies technicians in a field already facing shortages \cite{burton2018we}. The inherent subjectivity in visual examination also contributes to variability in clinical interpretation based on the EEG reader’s level of expertise. The presence of signal artifacts further complicates the reader's ability to accurately identify key bio-markers. The time-consuming nature of clinical EEG diagnosis and its variability could be greatly improved with an automated seizure classification system that assists professionals.

\subsection{Machine Learning Based Methods}
The uncertainty inherent in contemporary seizure diagnostic tools motivated this study’s investigation into the use of novel deep learning algorithms for seizure diagnosis. Recent work has demonstrated the potential of using classical machine learning techniques and deep learning frameworks on related EEG and epilepsy problems. The authors in \cite{fergus2016machine} applied the $K$-Nearest Neighbors (KNN) classifier for seizure detection, achieving $93$\% for area under the curve. \cite{shoeb2010application} performed detection of epileptic seizures using Support Vector Machine on scalp-EEG data, achieving $96$\% accuracy. The study in {\cite{golmohammadi2018deep}} demonstrated the viability of applying deep learning algorithms to automatic seizure detection, achieving $94$\% specificity using deep residual learning. For the more challenging task of seizure prediction, \cite{truong2018convolutional} applied Convolutional Neural Networks (CNNs) on frequency domain data obtained through a Short-Time Fourier Transform to achieve a sensitivity of $81.4$\% on patient-specific seizure prediction task. 

The aforementioned techniques mainly focus on the problems of seizure detection, which is defined as identifying a seizure as it occurs, and seizure prediction, which aims to predict seizures in advance. This study seeks to augment automated seizure advisory systems by introducing seizure type classification. Specific to the problem of seizure type classification, \cite{pereira2013icd9} applied text mining on electronic patient records for a 3-class classification task, achieving a F1-score of $71.05$\%. \cite{kassahun2014automatic} utilized ontology-based and genetics-based machine learning algorithms on ictal symptoms for a $2$-class seizure type classification task, with both algorithms achieving $77.8$\% accuracy. Recently, \cite{roy2019machine} applied KNN and Extreme Gradient Boosting to sEEG to classify between$7$-classes of seizures, achieving F1-scores of $90.7$\% and $70.7$\% respectively. The above mentioned methods can all be viewed as variations of classical machine learning algorithms, while \cite{asif2019seizurenet} proposed SeizureNet, an ensemble architecture composed of three DenseNets and extracts features from sEEG, achieving F1 of $98.4$\% on $7$-class classification problem. The studies {\cite{ahmedt2019neural}} and {\cite{sriraam2019convolutional}} applied Neural Memory Networks and VGG-$19$ based models, respectively, to sEEG data to classify between $8$ seizure classes.

Apart from the CNN models, the classical machine learning algorithms described above are limited by the potential to recognize and capture complex patterns in multi-channel EEG signal. These classical techniques are also limited in ability to exploit hierarchical structures in natural signals and learning from raw data without a priori feature selection \cite{schirrmeister2017deep}. Recently, deep learning based methods \cite{langkvist2014review, truong2018convolutional} have shown to outperform classical methods for learning more discriminative features from the EEG data compared to the hand crafted features.

This study is motivated by the potential of applying deep learning architectures to the problem of seizure type classification. Experiments are conducted using CNNs, RNNs and Bilinear Networks to uncover complex patterns from multidimensional EEG signals and perform this classification task. The main contribution of this paper is in demonstrating the effectiveness of second-order statistics of CNN and RNN features and bilinear pooling in diagnostic classification tasks. Additionally, our study demonstrates that hybrid dual-stream architecture outperforms symmetric bilinear models by exploiting the interactions of two explicit feature types. Lastly, the predictive performance of the proposed method and the generalization ability establish new benchmarks in seizure type classification.

\section{Proposed Method}
\subsection{Dataset}
Table \ref{tab:datasummary} summarizes the datasets being used in this work: the Temple University Hospital (TUH) Seizure Corpus dataset v1.4.0 and the EPILEPSIAE dataset. The TUH dataset contains surface EEG data of $817$ sessions, of which $305$ are sessions that contain seizures, resulting in a total of $2012$ seizure events recorded \cite{shah2018temple}. There are eight classes of seizures identified, the total number of seizure per type and the total recording time of each seizure is shown in Table \ref{tab:sztypesummary}. The TUH dataset consists of EEG recordings of various sampling rates, ranging from $250$ Hz to $512$ Hz. Recordings consist of $24$ to $36$ channels of signal data as well as an annotation channel.

\begin{table}[!h]
\centering
 \caption{Summary of datasets}

  \resizebox{0.8\columnwidth}{!}{
  \begin{tabular}{lcc}
  \toprule
    
    \multirow{1}{8em}{\textbf{Dataset features}} & \multirow{1}{3em}{\textbf{TUH}} & \multirow{1}{6em}{\textbf{EPILEPSIAE}}\\
    \midrule
    No. of patients (sEEG) & 314 & 30 \\
    No. of seizure recordings & 2012 & 276\\
    No. of seizure classes & 8 & 4 \\
    \bottomrule
  \end{tabular} }
  \label{tab:datasummary}
  \\[\bigskipamount]
\resizebox{1\columnwidth}{!}{
  \begin{tabular}{llcc}
  \toprule
    \multirow{2}{3em}{\textbf{Dataset}} & \multirow{2}{4em}{\textbf{Seizure type}} & \multirow{2}{6.5em}{\centering \textbf{No. of seizures events}} & \multirow{2}{7em}{\textbf{Total recording time (minutes)}}\\
    \\
    \midrule
    \multirow{9}{5em}{TUH} & Focal Non-Specific & 992 & 1224 \\
    & Generalised Non-Specific & 415 & 567\\
    & Simple Partial & 44 & 26 \\
    & Complex Partial & 342 &540 \\
    & Absence & 99 & 14\\
    & Tonic & 67 & 21\\
    & Tonic Clonic & 50 & 80\\
    & Myonic & 3 & 22\\
    \midrule
    & \multicolumn{1}{r}{\textbf{Total} }& \multicolumn{1}{r}{\textbf{2012}} & \multicolumn{1}{r}{\textbf{2494}} \\
    \midrule
    \multirow{5}{5em}{EPILEPSIAE} & Complex Partial & 139 & 186\\
    & Unclassified & 66 & 50\\
    & Simple Partial & 49 & 56\\
    & Secondarily Generalised & 22 & 57\\
    \midrule
    & \multicolumn{1}{r}{\textbf{Total} }& \multicolumn{1}{r}{\textbf{276}} & \multicolumn{1}{r}{\textbf{349}} \\
    \bottomrule
  \end{tabular}}
  \label{tab:sztypesummary}
\end{table}
 
\begin{figure*}[htp]
    \centering
    \includegraphics[width=14cm]{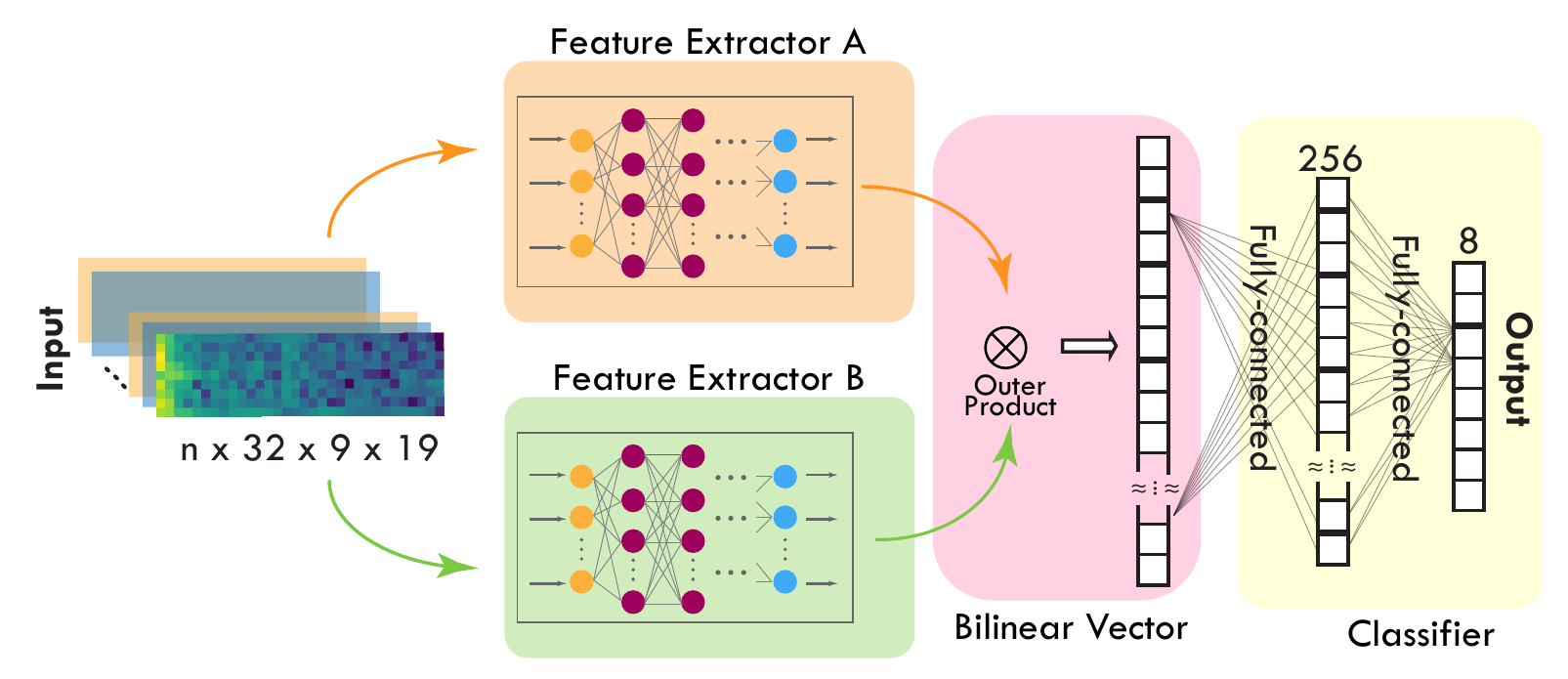}
    \caption{Bilinear Network}
    \label{fig:bilinear}
\end{figure*}

The EPILEPSIAE dataset was also employed in this study, for the sole purpose to test the generalization ability of our models. The dataset contains sEEG recordings from 30 patients, resulting in $276$ seizure events recorded \cite{ihle2012epilepsiae}. Identified epileptic seizures are classified into four types, with total number of seizures per type shown in Table \ref{tab:sztypesummary}. All recordings have a sampling rate of $256$ Hz and contain $19$ channels.

\subsection{Pre-processing}
As the TUH data contains various sampling rates, all samples were re-sampled to $250$ Hz to ensure uniform input dimensions to the neural network. $19$ channels common to all recordings were selected and rearranged based on the proximity of electrodes in the $10$--$20$ layout of scalp EEG electrodes. 

This study considers the use of Short-Time Fourier Transform (STFT) to exploit the frequency domain representation of raw EEG signal. In the  proposed feature engineering method, an STFT is performed on $1$-second samples with $50$\% window overlap using a $64$-point FFT cosine analysis window. The intensity values were then calculated by taking $log_{10}$. After this procedure, the dimension of each training sample becomes $(32, 9, 19)$ where $32$ is the number of frequency points, $9$ is the number of time steps and $19$ is the number of EEG channels. 

 Following the onset of most seizures, a subset of EEG channels develops rhythmic activity that is typically composed of multiple frequency components \cite{shoeb2010application}. The spectral structure and manifestations across multiple channels and across time are thus important in characterizing features of EEG data. As the STFT examines changes in frequency and phase information over time, it can effectively capture the time-varying spectral structures of epileptic EEG signals.

\subsection{Bilinear Models}
This study proposes the use of bilinear models for the purpose of seizure-type classification. Bilinear models have been shown to be exceptionally effective at fine-grained recognition and differentiating between similar objects (e.g. classification of different dog breeds)~\cite{lin2015bilinear}. The bilinear model consists of two feature extractors whose outputs are multiplied using matrix outer product at each location, then pooled to obtain a high level descriptor. This architecture can model local pairwise feature interactions as well as localize discriminative parts in a translationally invariant manner. The detailed topology of bilinear model is shown in Fig.~\ref{fig:bilinear}. In this section, we discuss why CNN and RNN are suitable feature extraction models for our bilinear architecture. In our experiments, these models are pre-trained on the same dataset, and then extracted and used in the bilinear structure. This architecture is modular as the {\lq}feature extractor{\rq} in Fig.~\ref{fig:bilinear} can be replaced interchangeably with either CNN or RNN models. As the EEG patterns of certain epileptic seizure classes, particularly focal and generalized seizures, share an overlap of EEG artifacts, bilinear models would effectively discriminate these events.

This study considers the use of symmetric and hybrid bilinear model. Symmetric networks are initialized with identical base feature extractors. This paper experiments with bilinear-CNN models (B-CNN) composed of identical pre-trained CNN blocks and bilinear-RNN models (B-RNN), composed of pre-trained Convolutional-LSTM (ConvLSTM) blocks. Hybrid bilinear models divorce this symmetry by employing different feature extractors. The hybrid model employs both CNN and RNN as feature extractors, where one network (ConvLSTM) extracts temporal features and the other (CNN) models spatial features, which are combined into second-order statistics through bilinear pooling. 

\begin{figure*}
    \centering
    \begin{subfigure}{0.8\textwidth}
        \centering
        \includegraphics[width=\linewidth]{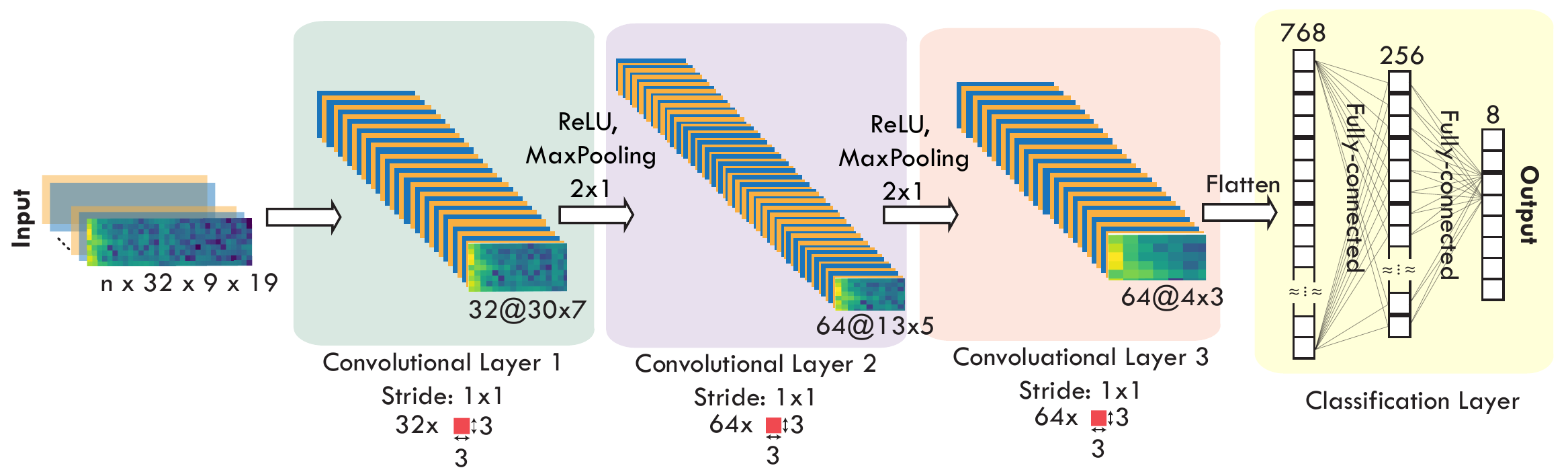}
        \caption{CNN Network}
        \label{fig:cnn}
    \end{subfigure}
    \begin{subfigure}{0.8\textwidth}
        \centering
        \includegraphics[width=\linewidth]{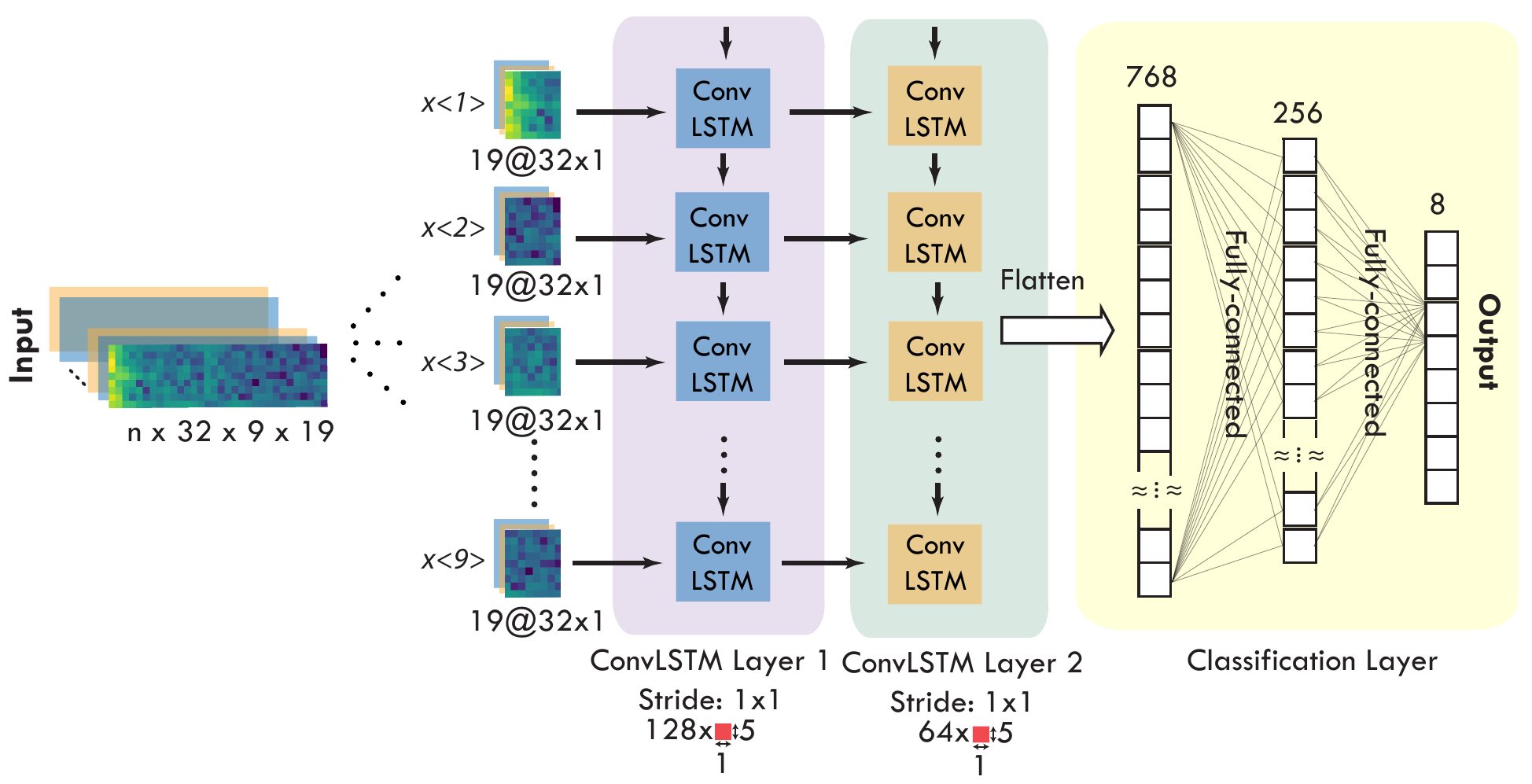}
        \caption{RNN Network}
        \label{fig:rnn}
    \end{subfigure}
    \caption{Base Feature Extractors}
\end{figure*}

In the bilinear architecture, the outputs of base feature extractor (i.e. the output from the last convolution or ConvLSTM layer) are multiplied using outer product to derive bilinear features for each location in the output. The two feature extractors extracts features of size $O \times M$ and $O \times N$ respectively, where $O$ is the output dimensions ($width \times height$), and $M$ and $N$ are the feature dimensions for each location. The outer matrix product of the two feature vectors ($M \times 1$ and $N \times 1$) for each location results in a bilinear feature of size $(M \times N)$. For the outer product operation to be compatible between the two extracted features, both feature outputs must have matching output dimensions $O$. Special care is taken while designing the networks to ensure the output sizes match. Calculating the outer product over each location thus produces $O - (M \times N)$ bilinear features. Bilinear features across all $O$ locations are aggregated through sum pooling (bilinear pooling) to achieve a final output size of $M \times N$. This matrix is then reshaped into a bilinear vector of size $MN \times 1$, normalized and fed into a fully-connected classifier.

Both the CNN and ConvLSTM proposed in this study produce features of output dimension $12$ and a $64$ dimension feature at each location. The bilinear operation and pooling thus results in a bilinear vector of dimensions $(64 \times 64) \times 1$. Predictions are obtained through the last layer, a softmax activation layer, with each node corresponding to the probability that the input signal belongs to a particular seizure class. The following sections will explore the base models in more detail and highlight the two-step training procedure employed.

\subsubsection{Convolutional Neural Network}
Convolutional Neural Networks (CNN) have been successfully applied to many aspects of biomedical research based on physiological signals and medical image analysis (see \cite{truong2018convolutional}, \cite{sriraam2019convolutional}). The CNN employed in this study includes two components, a feature extractor comprising three convolutional blocks and a fully-connected dense classifier. The topology of the CNN used in this study is highlighted in Fig.~\ref{fig:cnn}.

\subsubsection{Recurrent Neural Network}
Recurrent Neural Networks (RNN) were designed to work with sequence prediction problems, with Long-Short Term Memory (LSTM) models demonstrating significant promise in time-series classification tasks. This study uses a variation of LSTM networks, the Convolutional LSTM (ConvLSTM), first proposed by \cite{xingjian2015convolutional}. ConvLSTM replaces matrix multiplication performed in traditional LSTM cells with convolution operations, leveraging parameter sharing and sparsity of connection of data. The RNN employed in this study is composed of two components, a feature extractor composed of two ConvLSTM layers and a fully-connected dense classifier. The topology used in this study is shown in detail in Fig.~\ref{fig:rnn}.

\begin{table*}[htp]
 \caption{Comparison with other machine learning based classification methods. \protect\footnotemark}
  \centering
  \resizebox{1\textwidth}{!}{
  \begin{tabular}{lllll}
    \toprule
    Method & Dataset & Data Type & Seizure classes & Performance \\
    \midrule
    Support Vector Machine \cite{saputro2019seizure} & TUH & sEEG & \multirow{2}{24em}{3 - focal non-specific, generalized non-specific, tonic clonic} & Accuracy: 91.4\% \\ \\
    $K$-Nearest Neighbors \cite{roy2019machine} & TUH  & sEEG & \multirow{3}{24em}{7 - focal non-specific, generalized non-specific, simple-partial, complex-partial, absence, tonic, tonic-clonic} & F1: 90.70\% \\
    \\ \\
    Extreme Gradient Boosting \cite{roy2019machine} & TUH & sEEG & As above & F1: 70.70\% \\
    Ensemble CNN's \mbox{\cite{asif2019seizurenet}} & TUH & sEEG & As above & F1: 98.40\% \\
    CNN \mbox{\cite{sriraam2019convolutional}} & TUH & sEEG & \multirow{3}{24em}{8 - focal non-specific, generalized non-specific, simple partial, complex partial, absence, tonic, tonic clonic, myoclonic} & Accuracy: 84.06\% \\ \\ \\
    Neural Memory Networks \mbox{\cite{ahmedt2019neural}} & TUH & sEEG & As above & F1: 94.50\% \\ \\ \\
    \textbf{Hybrid Bilinear (this work)} & TUH & sEEG & As above & \textbf{F1: 97.40}\% \\
    \textbf{Hybrid Bilinear (this work)} & EPILEPSIAE & sEEG & \multirow{2}{24em}{4 - unclassified, simple partial, complex partial, secondarily generalized} & \textbf{F1: 97.00}\% \\ \\
    Text Mining \cite{pereira2013icd9} & Private Dataset$^*$ & Text & \multirow{2}{24em}{3 - complex focal seizure, simple focal seizure, generalized convulsive} & F1: 71.05\%     \\ \\
    Ontology-Based Algorithm \cite{kassahun2014automatic} & Private Dataset$^\dagger$ & Text & 2 - temporal lobe, extra-temporal lobe & Accuracy: 77.80\%    \\
    Genetics-Based Algorithm \cite{kassahun2014automatic} &  Private Dataset$^\dagger$ & Text & As above & Accuracy: 77.80\%    \\
    CNN \mbox{\cite{ahmedt2018deep}} & Private Dataset$^\ddagger$ & Video & \multirow{1}{24em}{2 - mesial temporal lobe, extra-temporal lobe} & Accuracy: 92.1\%    \\
    \bottomrule
  \end{tabular}}
  {\scriptsize
  $^*$: Leiria-Pombal Hospital Dataset,~
  $^\dagger$: Carlo Besta Neurological Institute, Niguarda Ca Granda Hospital and San Paolo Hospital Dataset,~
  $^\ddagger$: Brisbane Mater Hospital Dataset}
  \label{tab:benchmark}
\end{table*}

\subsection {Training and Evaluation}
The TUH dataset suffers from uneven class distribution (see Table~\ref{tab:sztypesummary}) and the class with the lowest count, absence seizures, has $14$ minutes of recording. The EPILEPSIAE dataset is relatively more even but still suffers from class imbalance. Under these circumstances, accuracy alone will be unlikely to select the best performing model. The F1-score is thus used as the metric to evaluate the performance of our trained models \cite{goutte2005probabilistic}. 

To tackle class imbalance problem, weights of the classes are incorporated into the training of the classifier (i.e. giving higher weighting to minority classes and lower weights to majority classes). Early stopping is employed as a regularization technique to reduce over-fitting during the training process. The method monitors validation loss and terminates the training process if validation loss does not improve over 10 epochs.

Stratified five-fold cross-validation is used to robustly evaluate the performance of proposed algorithms. The dataset is randomly split into five-folds, where each fold maintains the proportional distribution of classes in the entire dataset. During training, the model is estimated on four folds (training set) and evaluated on the remaining fifth fold (validation set). The overall performance of the model is based on the average validation F1-score over all folds.

The neural networks (CNN and RNN) are trained with the Adam optimization algorithm with batch sizes of $32$ and over $200$ epochs. The bilinear models rely on pre-trained feature extractors. The training and evaluation process for bilinear models are illustrated in Algorithm~\ref{alg:bilinear_training}. In each fold, the CNN and RNN are first trained on the training set. Subsequently, the feature extraction layers, namely convolutional and recurrent layers are inserted with learned weights into the bilinear networks. The bilinear models are then trained on the same training set using a two-step process (lines $7$ and $8$ of Algorithm \ref{alg:bilinear_training}). Firstly, only the bilinear pooling and dense layers were trained for $50$ epochs with early stopping. After which, the entire model, including the pre-trained feature extract layers, is fine-tuned through backpropagation through $100$ epochs with early stopping.

As both the outer product and sum-pooling are differentiable matrix operations and the bilinear model remains a directed acyclic graph and parameters can be trained in an end-to-end fashion using backpropagation. The details of gradient propagation through bilinear layers is clarified in work \cite{lin2015bilinear}. Additionally, the bilinear pooling operation creates bilinear vector of large dimensions, significantly increasing the number of trainable parameters. By loading pre-trained layers and employing the two-step training procedure, training time is reduced significantly.

It is important to note that the proposed architectures for each respective neural network topology was trained and fine-tuned solely on the TUH dataset. The models, without any adjustments to their architectures, were then trained on the EPILEPSIAE dataset. This was done to evaluate the generalization ability of our algorithms. 

\footnotetext{Only the best performing model was selected from each study.}

\begin{algorithm}[!h]
\SetAlgoLined
Create five stratified folds from dataset\;
 \For {fold $k_{i}$ in 5 folds}{
assign fold \textit{k} to test set\;
assign remaining folds to training set\;
train CNN, RNN on training set\;
extract trained feature extractor layers with learned weights from CNN, RNN and insert into bilinear network\;
train bilinear and classification layers of bilinear network on training set\;
fine-tune the entire model\;
evaluate F1 on test set\;}
calculate average F1-score over all folds\;
\caption{Bilinear Model Training and Evaluation}
\label{alg:bilinear_training}
\end{algorithm}

\section{Results}
In this section, we test our approach on both datasets. All models were implemented using Python~$3.5$ and Keras~$2.0$ with the Tensorflow~$1.4.0$ back-end. The models were run on a NVIDIA~K$80$ graphic cards, with each training epoch completing in roughly $100$ seconds. Five-fold cross-validation was performed and the average F1-scores were reported. Table~\ref{tab:results} summarizes seizure type classification results on the two datasets. On the TUH dataset, the CNN and RNN alone do remarkably well, achieving F1-score of $95.50$\% and $95.80$\% on the STFT data, respectively. The symmetric bilinear models further improved the classification performance by achieving $96.70$\% and $96.90$\% and the hybrid model achieves the best performance with F1-score of $97.40$\%.

\begin{table}[!h]
 \caption{Summary of epileptic seizure type classification results (F1-score \%).}
  \centering
\resizebox{1\columnwidth}{!}{
  \begin{tabular}{llccccc}
    %\cmidrule(r){1-7}
    \toprule
    \multirow{1}{1em}{Dataset} & \multirow{2}{3.2em}{Pre-processing} & \multirow{1}{2em}{CNN} & \multirow{1}{2em}{RNN} & \multirow{1}{3.2em}{B-CNN} & \multirow{1}{3.2em}{B-RNN} & \multirow{1}{2.8em}{Hybrid} \\ \\
    \midrule
    TUH & STFT & 95.5 & 95.8 & 96.7 & 96.9 & \textbf{97.4} \\
    EPILEPSIAE & STFT & 87.3 & 89.0 & 93.7 & 94.9 & \textbf{97.0}\\
    \bottomrule
  \end{tabular}}
  \label{tab:results}
\end{table}

The STFT based inputs and our proposed algorithms achieved similar levels of performance on the EPILEPSIAE dataset, achieving  F1-score of $87.3$\% and $89.0$\% on the base CNN and RNN models, respectively. The B-CNN model achieved $93.7$\% while the B-RNN achieved $94.9$\%. Once again, the hybrid model achieves the best performance of $97.0$\%. It is important to note that our approach works comparably on two separate datasets without any modifications of the algorithm for the EPILEPSIAE dataset.

Table~\ref{tab:benchmark} describes a benchmark of recent seizure classification approaches. In some cases, it is challenging to compare different approaches as each method is tested on a different dataset with differing number of seizure classes. In the domain of deep learning based methods, the authors in {\cite{asif2019seizurenet}} achieved an average F1-score of $98.4$\% on a an ensemble of three DenseNet-based CNN's trained on the TUH dataset. This proposed architecture contains $45.94$ million parameters, compared to $1.2$ million parameters in the hybrid bilinear structure proposed in this study. This work also omitted myoclonic seizures from the classification task due to the low numbers of samples of myoclonic class in the TUH dataset. Both {\cite{ahmedt2018deep}} and {\cite{sriraam2019convolutional}} proposed solutions to the $8$-class classification problem on the TUH dataset, achieving $84.06$\% and $94.05$\%. As can be seen from Table {~\ref{tab:results}}, the bilinear models proposed in this study achieve better performance on the $8$-class classification problem. Non-deep learning methods, including the KNN proposed in {\cite{roy2019machine}} and the Support Vector Machine proposed in {\cite{saputro2019seizure}}, demonstrate reasonable performance ($90.7$\% and $91.4$\%) but were achieved through extensive feature engineering that is not desirable. Our algorithms demonstrated very high classification accuracy with minimum feature engineering. We also demonstrated that our algorithms do not overfit to a specific dataset and can generalize well to other datasets.

\begin{figure*}
    \centering
    \begin{subfigure}{.6\textwidth}
        \centering
        \includegraphics[width=\linewidth]{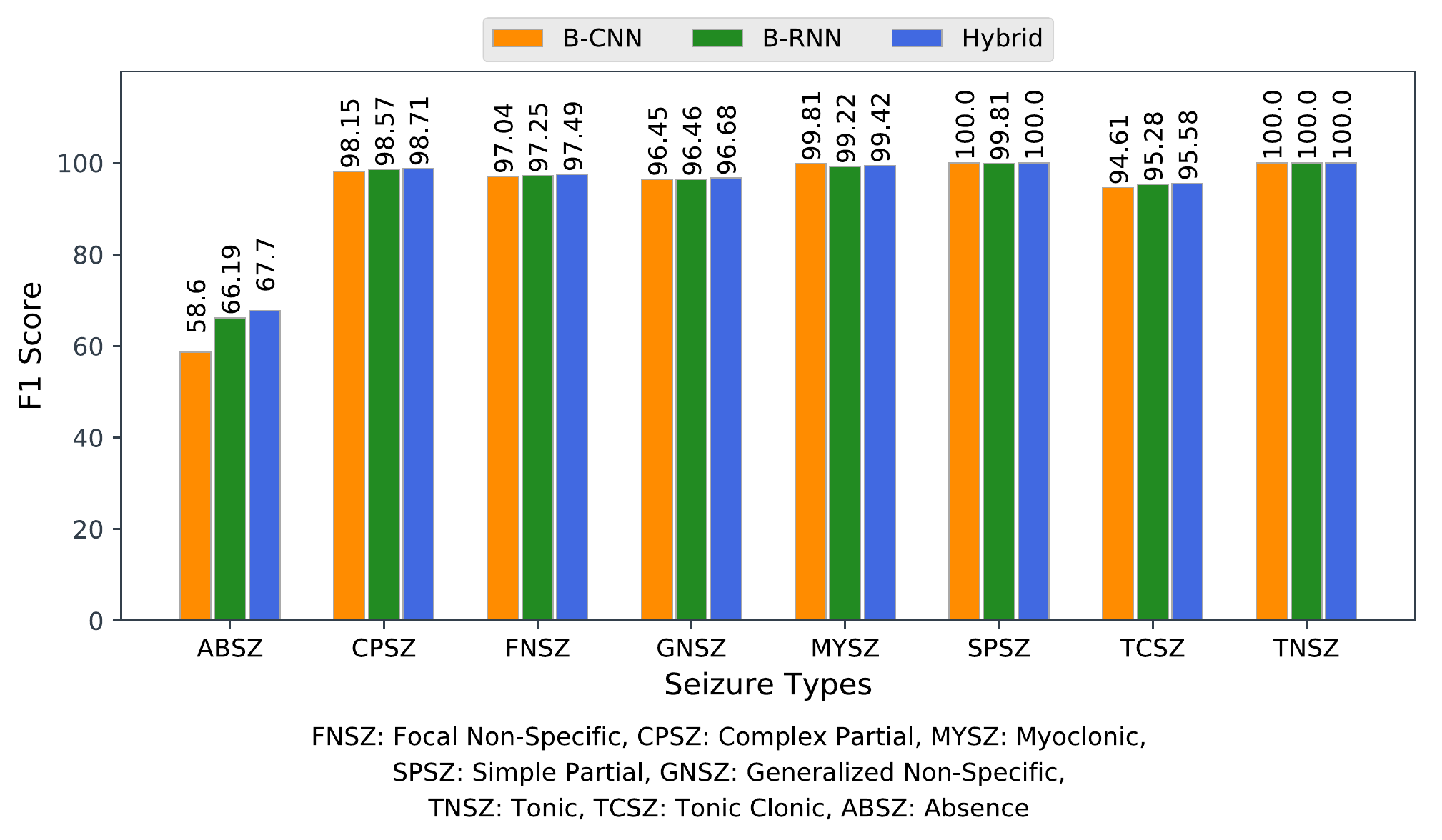}
        \caption{Classification accuracy on the TUH dataset.}
        \label{fig:tuh_result}
    \end{subfigure}%
    \begin{subfigure}{.4\textwidth}
        \centering
        \includegraphics[width=0.9\linewidth]{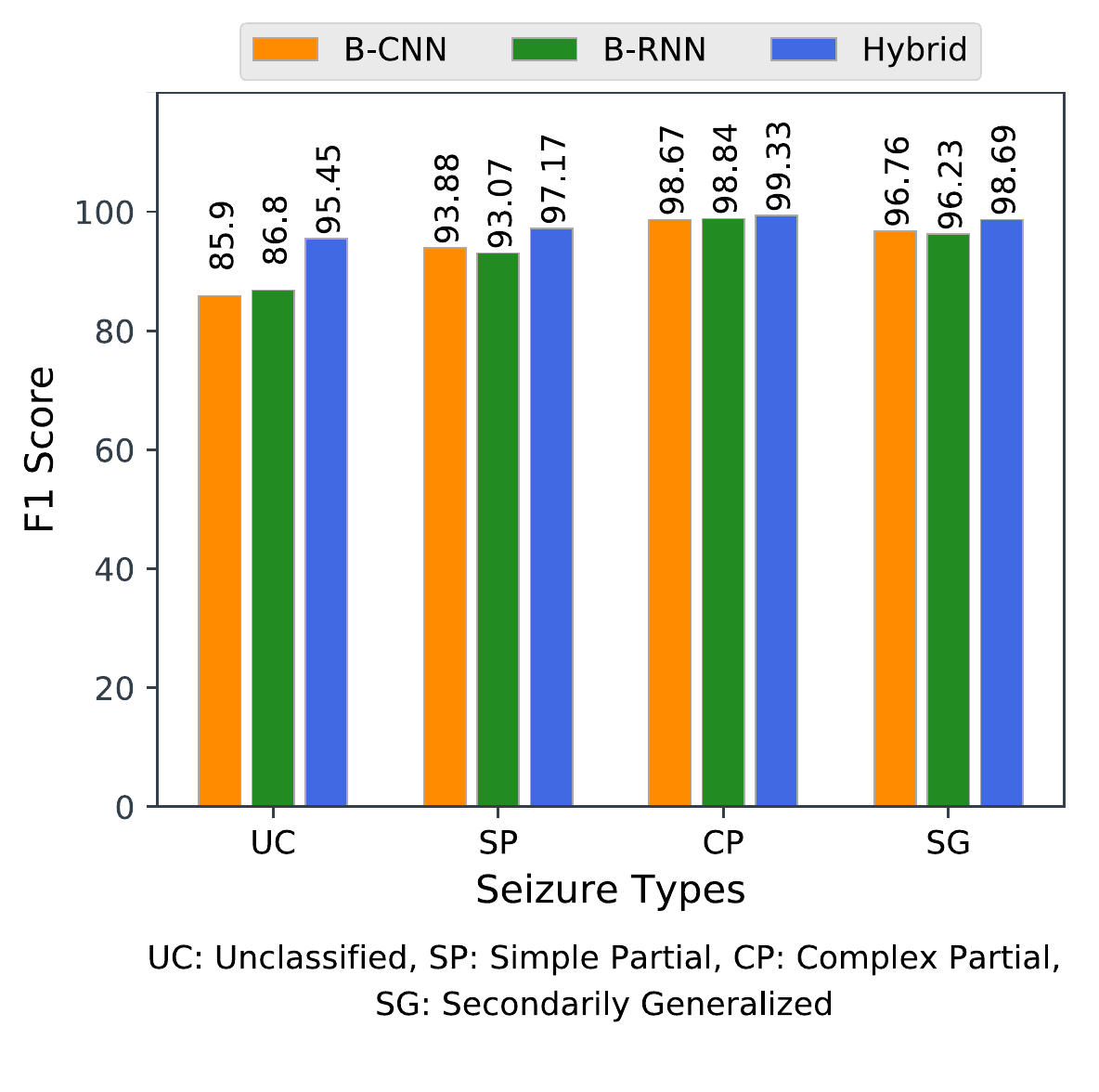}
        \caption{Classification accuracy on the EPILEPSIAE dataset.}
        \label{fig:epi_result}
    \end{subfigure}
    \caption{Classification over seizure classes.} \label{fig:result_per_class}
\end{figure*}

\begin{figure*}
    \centering
    \begin{subfigure}{.48\textwidth}
        \includegraphics[width=1\linewidth]{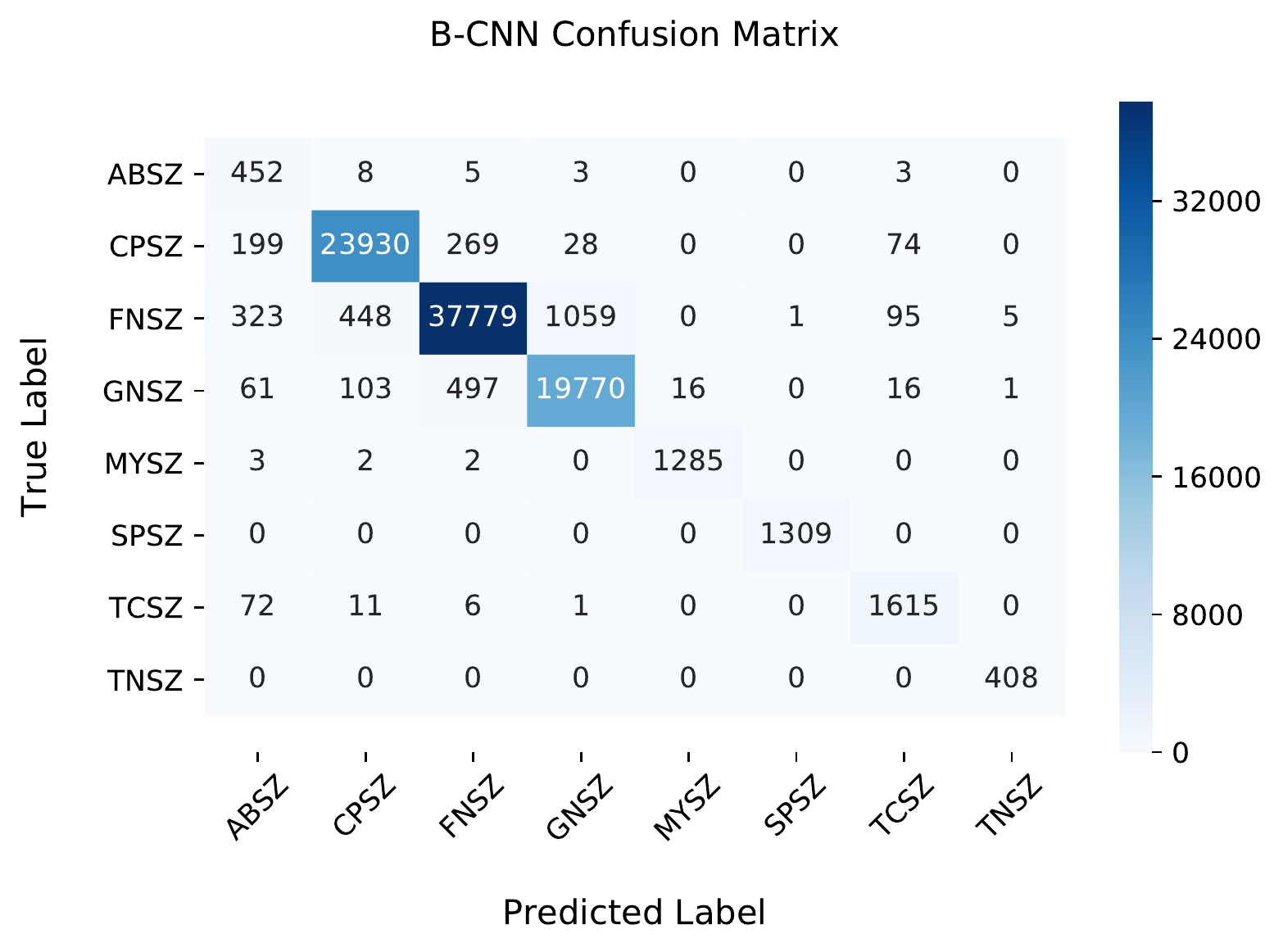}
        \hspace{0.1cm}
    \end{subfigure}%
    \begin{subfigure}{.48\textwidth}
        \hspace{0.1cm}    
        \includegraphics[width=1\linewidth]{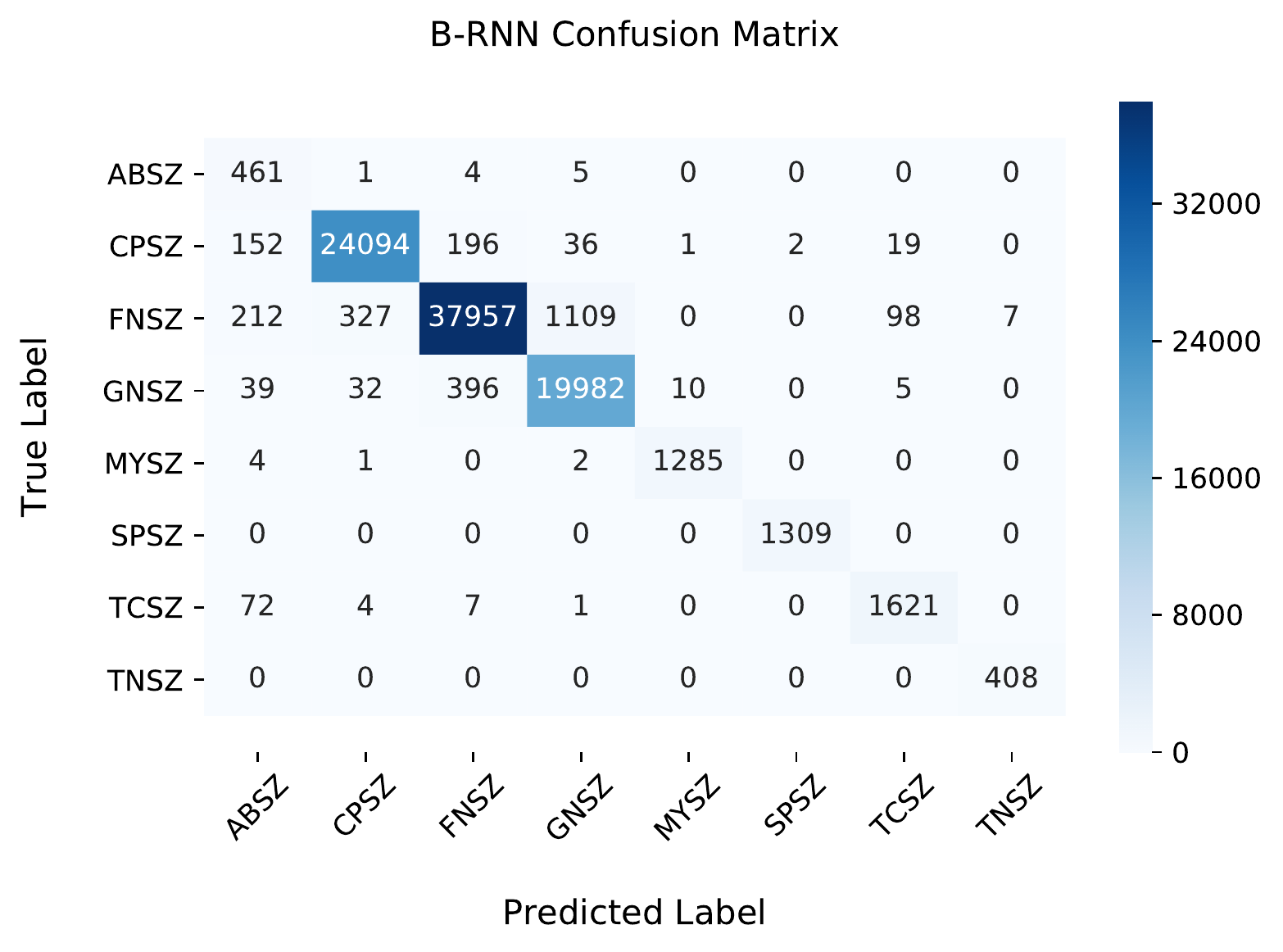}
    \end{subfigure}
    
    \begin{subfigure}{.48\textwidth}
        \vspace{0.2cm}
        \includegraphics[width=1\linewidth]{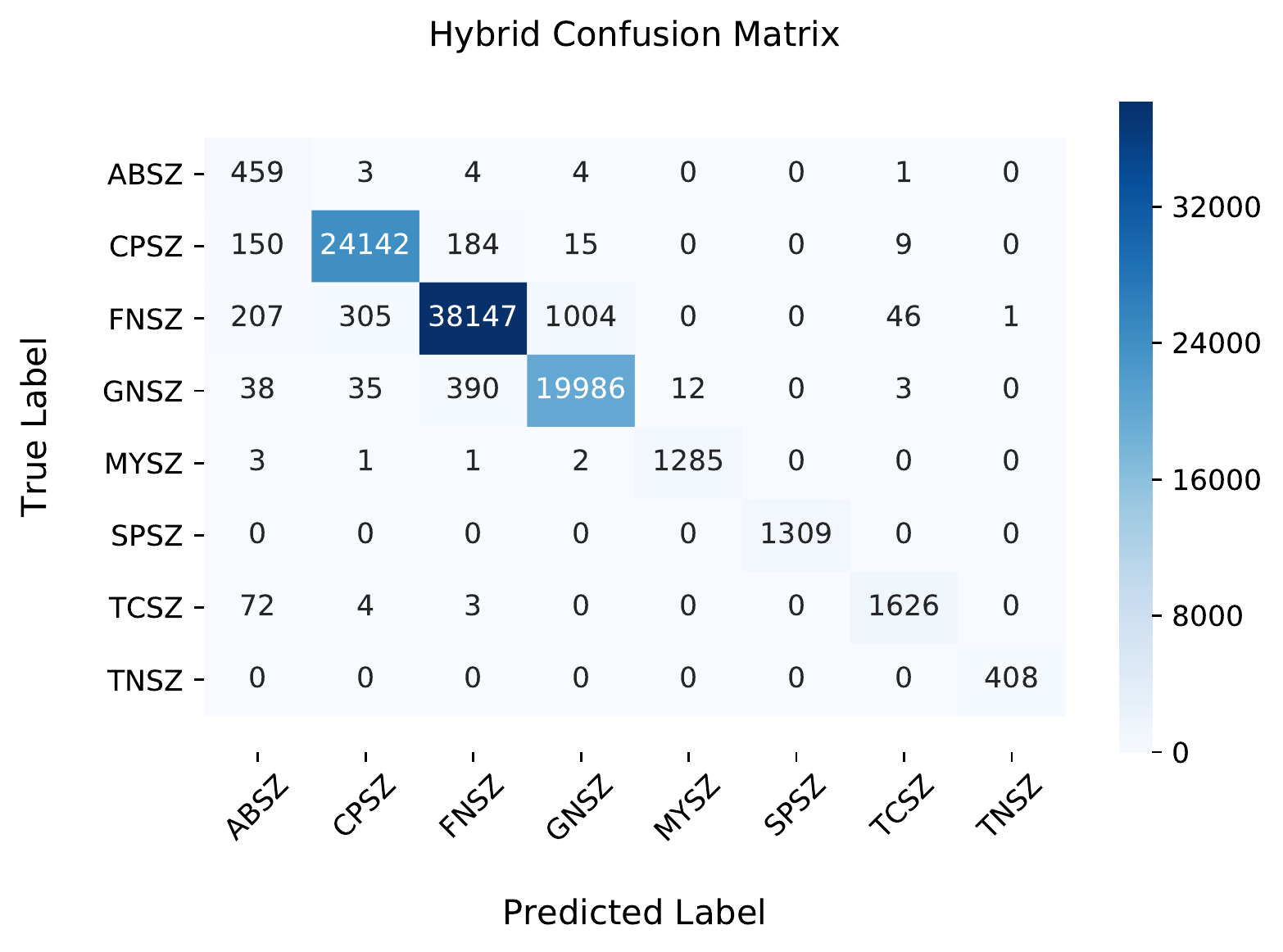}
    \end{subfigure}
    \caption{Confusion matrices of seizure type classification for the symmetric and hybrid bilinear models testing with the TUH dataset.} \label{fig:confusion}
\end{figure*}

\section{Discussion}
Figs.~\mbox{\ref{fig:result_per_class}} and \mbox{\ref{fig:confusion}} highlight the classification performance of the bilinear algorithms on each seizure class over the two datasets. For the TUH dataset, classification performance across different seizure types is generally comparable, with the exception of absence seizures (ABSZ). This can be attributed to the low count of absence seizures in the TUH dataset, with only $14$ minutes worth of recording for the algorithms to learn from. On the EPILEPSIAE dataset, the bilinear models do slightly worse on unclassified (UC) seizures. Like absence seizures, unclassified seizures have the shortest recording time in the EPILEPSIAE dataset. Additionally, a number of different seizure activities and difficult EEG patterns could have been generally labelled as {\lq}unclassified{\rq}, making it hard to make reliable predictions for that class. It should also be noted that hybrid bilinear networks perform significantly better than symmetric network when predicting on these difficult classes. 

Information extracted from EEG signals in frequency and time domains has been widely used in biomedical classification tasks. The performance of models trained on FFT-processed inputs is similar to those achieved using STFT. However, the FFT proposed by \cite{roy2019machine} in a preceding study contains several flaws. Firstly, the FFT method forfeits valuable temporal information that is critical to highly non-stationary EEG signals. Secondly, the FFT truncates the spectral information to the first $24$ frequency bands. While this is a common method of pre-processing to help traditional machine learning algorithms to learn, it is unnecessary in the context of deep learning, where algorithms can learn from more complex signals. Finally, it is not clear why the pre-processing yielded $20$ channels as there are only $19$ common recording EEG channels. 

The bilinear models compare favourably to existing benchmarks (Table~\ref{tab:benchmark}). However, the basic CNN and RNN already achieve very high classification performances. A possible explanation is that these networks were carefully designed and fine-tuned for the problem. We performed a Mann-Whitney U Test to compare the performance of the hybrid networks against symmetric architecture across 100 folds. The $p\textrm{-value} = 0.025$ indicates that the difference is statistically significant, demonstrating the ability for hybrid networks to outperform symmetric architectures. The hybrid network outperforms single stream networks by more than 2\%, which is also statistically significantly different than the best single stream network ($p\textrm{-value} = 0.022$). Furthermore, the CNN and RNN do not generalize well onto a different dataset, evident in the decreased performance on the EPILEPSIAE dataset. The advantages of the hybrid network is more salient on the EPILEPSIAE dataset, where hybrid architecture achieved significantly better performance over symmetric networks. The hybrid bilinear network achieves similar-level performance on the EPILEPSIAE and TUH datasets, highlighting the ability for the model to generalize. The EPILEPSIAE dataset is smaller with only $349$ minutes of recording compared to $2494$ minutes in the TUH dataset, demonstrating the ability for hybrid models to learn from smaller amount of data. 

The bilinear models work well as the bilinear vector obtained by multiplying the output of feature extractors can effectively model interactions between pairs of local features. Thus, they can discriminate well when the input data is similar and can thus differentiate similar EEG artifacts in different seizure classes. Intuitively, the hybrid bilinear model leverages the unique strengths of two different types of deep networks and considers all pairwise interactions between spatial and temporal features. As the bilinear architecture is modular, different feature extracting methods can be exchanged to achieve optimal performance. For example, hand-crafted feature extractors that are trained to recognize specific EEG artifacts can be combined to deliver improved performance.

The study presents a step towards automated seizure diagnostic tools that can assist neurologists in diagnosing epileptic seizure types. This tool can be extended to also detect and classify between epileptic and non-epileptic seizures, forming an integrated diagnostic tool that can greatly improve the speed, accuracy and reliability of epilepsy diagnosis. Research has also shown that alternative types of sensory data, including electrocardiogram (ECG), blood oxygen levels, temperature, glucose levels as well as patient variables such as age and gender all influence the manifestation of epilepsy \cite{Yao2017hydration}, \cite{Cogan2015szeda}, \cite{kwak2017flexibleheartrate}, \cite{di2018wearableoxygen}. The inconsistent performance of EEG-based models on edge cases (e.g. absence seizures) further highlights the need for multi-modal systems and future work can focus on integrating these additional sensory data to provide more accurate diagnosis. 

\section{Conclusion}
Classification of epileptic seizures has been a challenge for neurologists diagnosing epilepsy, prescribing treatment and arriving at a prognosis. The automated seizure classification system proposed in this study can assist clinical professionals in diagnosing the disease, reducing time and potentially improves accuracy and reliability. This paper proposes a novel hybrid bilinear model and demonstrates the capability of using these models with minimal feature engineering to classify seizures based on sEEG data. The hybrid bilinear architecture achieves a significant improvement in classification accuracy, achieving $97.4\%$ and $97.0\%$ on the TUH and EPILEPSIAE datasets, establishing new benchmarks in performance. The number of parameters in our models is just over one million with the inference time of around 1.2 ms on an NVIDIA K80 GPU. The model also generalizes well across different datasets and minority seizure types. Future works can focus on improving the performance of this study, by incorporating sensory data-fusion techniques using multi-modal data, primarily audiovisual.

\section{Acknowledgment}
Omid Kavehei would like to acknowledge funding provided via The University of Sydney SOAR Fellowship. Nhan Duy Truong would like to thank The University of Sydney Nano Institute for its partial financial support via a John Makepeace Bennett scholarship.

\bibliographystyle{IEEEtran}
\bibliography{manuscript.bib}

\vfill

\end{document}